\documentclass[english,aps,prb,twocolumn,showpacs,floatfix,superscriptaddress]{revtex4}

\usepackage{babel}
\usepackage[utf8]{inputenc}  
\usepackage[active]{srcltx}  

\usepackage{color}
\usepackage{epsfig}
\usepackage{graphicx}
\usepackage{pstricks}

\usepackage{amsmath}
\usepackage{amssymb}
\usepackage{amsfonts}
\usepackage{amsthm}
\usepackage{bm}

\makeatletter
\@ifundefined{textcolor}{}
{%
 \definecolor{BLACK}{gray}{0}
 \definecolor{gray}{gray}{0.5}
 \definecolor{WHITE}{gray}{1}
 \definecolor{RED}{rgb}{1,0,0}
 \definecolor{GREEN}{rgb}{0,1,0}
 \definecolor{BLUE}{rgb}{0,0,1}
 \definecolor{CYAN}{cmyk}{1,0,0,0}
 \definecolor{MAGENTA}{cmyk}{0,1,0,0}
 \definecolor{YELLOW}{cmyk}{0,0,1,0}
}

\makeatother

\begin{document}

\title{Structural transition, spontaneous formation of strong singlet dimers and metamagnetism in $S=3/2$ magnetoelastic spin chains}

\author{C.\ J.\ Gazza }

\affiliation{IFIR-CONICET and Facultad de Ciencias Exactas, Ingeniería y Agrimensura,
Universidad Nacional de Rosario, Argentina}

\author{A.\ A.\ Aligia}

\affiliation{Instituto de Nanociencia y Nanotecnolog\'{\i}a, CNEA-CONICET, GAIDI, Centro At\'{o}mico Bariloche, 8400 Bariloche, Argentina}

\author{A.\ O.\ Dobry}

\affiliation{IFIR-CONICET and Facultad de Ciencias Exactas, Ingeniería y Agrimensura,
Universidad Nacional de Rosario, Argentina}

\author{G.\ L.\ Rossini }

\affiliation{IFLySiB-CONICET and Departamento de F\'{i}sica, Universidad Nacional
de La Plata, 
Argentina}

\author{D.\ C.\ Cabra}

\affiliation{IFLySiB-CONICET and Departamento de F\'{i}sica, Universidad Nacional
de La Plata, 
Argentina}

\date{\today}

\begin{abstract}
We study a one-dimensional antiferromagnetic-elastic model with magnetic ions having spin $S=3/2$. 
By extensive DMRG computations and complementary analytical methods, 
we uncover a first-order transition from a homogeneous or weakly-dimerized phase 
(a situation that could be similar to the well known $S=1/2$ spin-Peierls effect) to a highly distorted phase, 
driven by the spin-phonon coupling $\lambda$. 
The striking characteristic of the second phase, present at large $\lambda$, 
is the appearance of weakly ferromagnetic (FM) couplings alternating with strong antiferromagnetic (AFM) ones 
(we dub it FM-AFM phase) 
with a ground state close to a direct-product state of singlet dimers 
sitting on the AFM bonds. 
The behavior of the spin gap in both phases is studied by DMRG computation 
and contrasted with bosonization predictions and perturbation theory around the direct product of dimers.
In the FM-AFM phase robust magnetization plateaus and metamagnetic jumps show under magnetic fields.

The novel phase could be realized in  5d oxides of current interest, with giant spin-phonon coupling. 
Potential applications of the transition would be associated to the possibility of tuning the 
transition by external parameters such as striction, magnetic or electric fields, or alloying.

\end{abstract}


\maketitle

\section{Introduction\label{sec:Introduction}}

It is by now well established that a one dimensional spin $S=1/2$ system coupled to the lattice has an instability towards 
dimerization in a wide range of the model parameters, the so-called spin-Peierls (SP) instability. 
Theoretical understanding [\onlinecite{Cross}] and numerical confirmation [\onlinecite{Feiguin_1997}] have been presented years ago. 
The phenomenon is robust against various details in the magnetic exchange [\onlinecite{Dobry-Riera_1995}]
and in the elastic coupling to the lattice [\onlinecite{Penc_2004,Balents_2006}]. 

This instability has strong consequences [\onlinecite{Pouget}] in materials such as 
 the (TMTTF)$_2$X Fabre salts and related salts with the o-DMTTF donors as well as in the inorganic material CuGeO$_3$ [\onlinecite{Hase}], where most 
of the theoretical predictions have been experimentally tested in different scenarios.  

Such kind of magneto-elastic instabilities  have also been studied 
in more general situations, 
{\it e.g.} in the presence of an external magnetic field [\onlinecite{Kiry},\onlinecite{Temo}], in quasi one-dimensional arrays such as spin ladders, 
in chains of magnetic ions with spin higher than 1/2 [\onlinecite{Guoetal}], etc. 
The spin-Peierls transition has been also thoroughly studied in two-dimensional systems, 
either bipartite as the square lattice or in frustrated ones [\onlinecite{2D-SP}].

Regarding the study of magneto-elastic spin chains with $S > 1/2$, there has been intense research 
in connection with the well-known Haldane conjecture [\onlinecite{Haldane_1983}] which states strong differences between integer 
or half-odd-integer $S$. 
Based on a low-energy field theory mapping  [\onlinecite{Schulz,Affleck-Haldane}] it was conjectured that the SP instability could also take place  in chains with half odd-integer spin greater than $S=1/2$. This has been further supported numerically [\onlinecite{Affleck_1989}] and theoretically using the non-Abelian bosonization mapping [\onlinecite{NAbos}].
But in spite of the theoretical efforts, numerical evidence is hardly conclusive for chains with spin higher than 1/2, partly due to the additional complexity
associated to the increasing dimension of the local Hilbert space. 
In an early work [\onlinecite{Guoetal}],  it was argued that for integer spin the SP instability would be generically absent, while 
it would survive for half odd-integer spin chains. 
These early results have been obtained in very short chains and the extrapolations are at most qualitative. As a consequence, 
the complete phase diagram has not been established so far, and the existence of a magneto-elastic instability for  $S>1/2$ is still controversial. 

More recently a SP-like  structural transition has been claimed to be relevant in the description of many
multiferroic systems such as some double perovskites, manganites, nickelates, etc. 
where  a simultaneous sudden increase of the magnetization along with an equally abrupt 
decrease of the electric polarization is experimentally observed [\onlinecite{m1,m2,m3,m4,m5}]. 
This phenomenon has been theoretically studied in low-dimensional models [\onlinecite{Pantografos,Pili}]
where it was shown that the interplay between magnetic and electric dipoles mediated by the elastic degrees of freedom
explains most of the experimentally observed features.

Motivated by the ubiquitous relevance of the SP transition in magneto-elastic systems, 
and the apparent validity of the Haldane conjecture, 
in the present paper we focus on the $S=3/2$ case.

Analyzing the model in absence of magnetic field for a wide range of the spin-phonon coupling $\lambda$,
aside from the elusive question of the existence of a SP instability for weak $\lambda$, 
we find a feature not present in the $S=1/2$ physics which is the main message in this work: 
a first-order structural transition as 
a function of the spin-phonon coupling from a homogeneous/weakly-dimerized  phase with antiferromagnetic exchanges at low coupling
(which would be consistent with the field theoretical expectations [\onlinecite{Schulz,Affleck-Haldane,NAbos}])
into a strongly dimerized phase with alternating ferro- and antiferro-magnetic exchange interactions (dubbed here as FM-AFM phase) realized at strong coupling. 
The two distinct regimes could be observed in different materials, according to their intrinsic spin-phonon coupling, but more interestingly
the transition could be driven  by a variety of parameters such as striction, magnetic or electric fields, etc. 

The uncovered FM-AFM phase is potentially relevant for multiferroic applications whenever a tailoring of $\lambda$ leads to a giant variation of the magnetic susceptibility. 
The presence of the first-order transition, and the associated large change in the crystal structure, 
could produce a noticeably drastic jump in both magnetization and electric polarization in improper type II multiferroic materials 
whenever half-odd-integer higher spins were involved.

The paper is organized as follows: 
in Section II we introduce the model, describe the numerical approach, 
and present clear evidence of the first-order transition from extensive DMRG computations. 
We also discuss the response of the model to external magnetic fields.
In Section III we discuss analytical support from perturbation theory applied to the decoupled dimers regime, 
which provides a valuable description of the strongly dimerized phase.
In Section IV we review the question of the low spin-phonon coupling regime 
and the existence of a spin-Peierls instability in the $S=3/2$ chain at the light of our numerical results. 
Section V provides a summary, discusses the findings, and outlines potential applications.

\section{The model and numerical results}

\subsection{Model Hamiltonian}
\label{model}

We consider a magneto-elastic spin chain with local spins $S=3/2$. 
The magnetic sector is described by 
nearest-neighbor antiferromagnetic exchange with Hamiltonian given by
\begin{eqnarray}
H_\text{spin}=\sum_i J_i \mathbf{S}_{i}.\mathbf{S}_{i+1} - h^z \sum_i S^z_{i} 
\label{Hmag}
\end{eqnarray}
where $\mathbf{S}_{i}$ is the magnetic moment of an ion at site $i$ ($1\leq i \leq N$ in a chain of $N$ sites and periodic boundary conditions) 
and $h^z$ is a homogeneous external magnetic field. 
A minimal coupling to the lattice is given by distortion modulated exchange couplings
\begin{eqnarray}
J_i = J \left[ 1-\alpha (u_{i+1}-u_{i} )  \right]
\label{J_j}
\end{eqnarray}
where $u_{i}$ describes the displacement of ion $i$ from its regular lattice position. 
That is, $J_i$ depends linearly on the bond distortion $\delta_i \equiv u_{i+1}-u_{i}$ 
through a (dimensionfull) spin-phonon coupling $\alpha$.

The lattice degrees of freedom will be described by their energy cost in the adiabatic limit, 
under the assumption that phonon frequencies are much smaller than $J$.
We choose here the so-called Einstein-site phonon (ESP) model [\onlinecite{Balents_2006}] which considers
a quadratic energy arising from displacement of magnetic ions from their equilibrium positions 
(in absence of magnetic interactions),
\begin{eqnarray}
H_\text{latt}=\frac{K}{2}\sum_i u_{i}^{2}
\label{J_i}
\end{eqnarray}
describing a dispersionless optical phonon branch.

It is convenient at this point to introduce dimensionless parameters which will be used in what follows. We define $\tilde{u}_i = \alpha \, u_i$
and $\tilde{h}^z = h^z/J$, setting the energy scale $J=1$. The model Hamiltonian $H=H_\text{spin}+H_\text{latt}$ then reads
\begin{eqnarray}
H/J=\sum_i \left[(1-\left(\tilde{u}_{i+1}-\tilde{u}_{i} )\right) \mathbf{S}_{i} \cdot \mathbf{S}_{i+1}
+\frac{1}{2\lambda}\tilde{u}_{i}^{2} -\tilde{h}^z S^z_{i}\right]
\label{Hadim}
\end{eqnarray}
where $\lambda =J\alpha^2/K$ is the already mentioned dimensionless spin-phonon coupling.
Notice that the length scale is given by $1/\alpha$, not by the lattice spacing $a$ frequently used elsewhere.

%


\subsection{Self consistent DMRG approach}

We follow a numerical approach to the full Hamiltonian in Eq. (\ref{Hadim})
based on a self-consistent adiabatic procedure to minimize
the total energy [\onlinecite{Feiguin_1997}]. 
Given an initial (classical) ion displacement configuration $u_i$ we apply 
DMRG methods to compute the  (quantum)
ground-state spin configuration.
We then generate a new  distortion configuration from the self-consistent equation
\begin{equation}
	\tilde{u}_i=\lambda \left( \langle \mathbf{S}_{i-1} \cdot \mathbf{S}_{i}\rangle -  \langle\mathbf{S}_{i} \cdot \mathbf{S}_{i+1}\rangle \right)
	\label{self-consistency}
\end{equation}
and iterate until we get a fixed point. 

DMRG  calculations have been implemented using the ITensor software library [\onlinecite{ITensor}], 
setting the necessary maximum bond dimension to keep a truncation error cutoff of $10^{-10}$. 
In all cases we have used periodic boundary conditions. 

\subsection{Numerical results}

\subsubsection{First-order transition in the absence of magnetic field}

\begin{figure}
    \centering
    \includegraphics[width=1.0\linewidth]{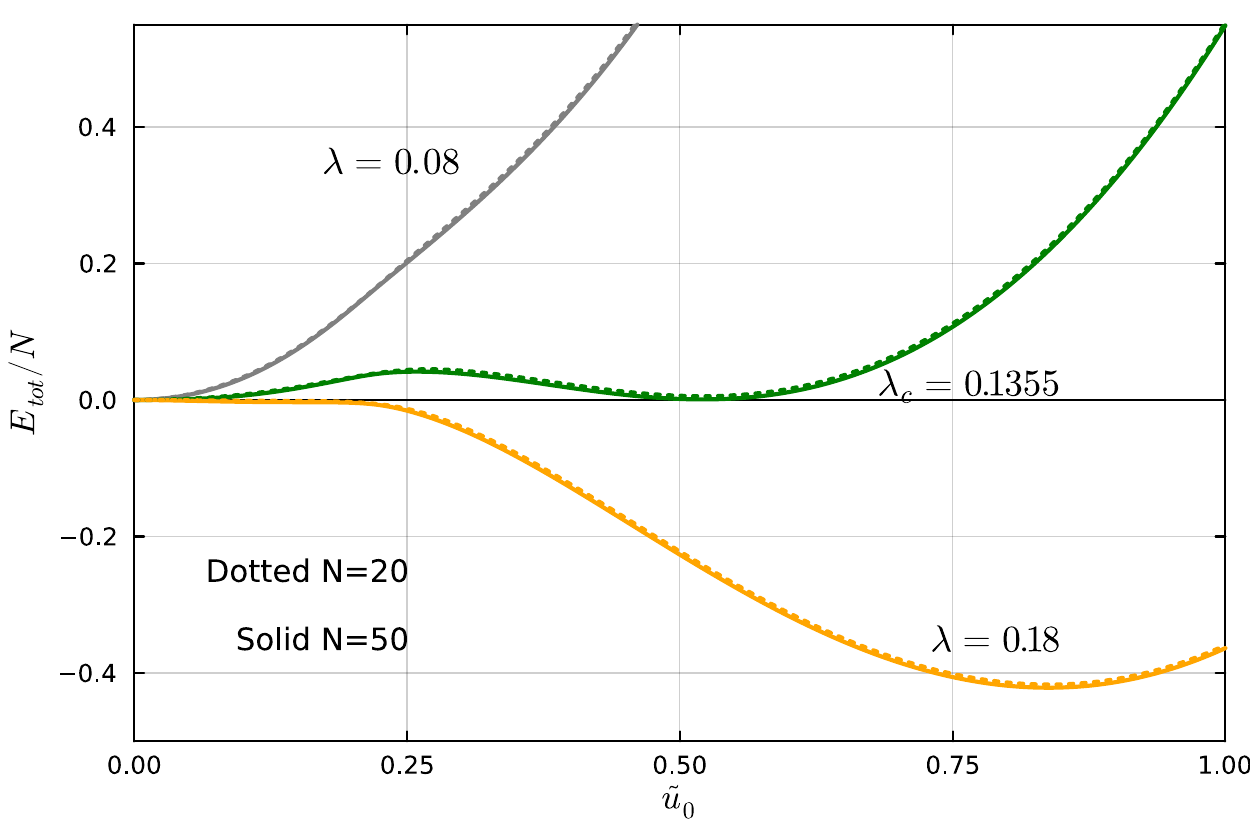}
    \caption{Total energy computed for alternating displacements as a function of their amplitude $\tilde{u}_0$
    for different values of the spin-phonon coupling $\lambda$ (data for $N=20$ and $N=50$ sites are shown).}
    \label{fig:energy-landscapes}
\end{figure}

Starting with $h^z=0$, for the full range of the spin-phonon coupling $\lambda$ we have found 
that the ground state is a spin singlet ($S^z_{tot}=0$) and exhibits a period 2 pattern of ion displacements, 
say $\tilde{u}_i=(-1)^i \tilde{u}_0$, producing an alternation of short and long bond distances.  
But depending on $\lambda$ we have found two strikingly different solutions, 
as can be seen in Fig.\ \ref{fig:energy-landscapes}, where we show 
the total (magnetic plus elastic) energy for dimerized distortions 
as a function of the distortion amplitude 
$\tilde{u}_0$
for different values of $\lambda$.

\begin{figure}
    \centering
    \includegraphics[width=1.0\linewidth]{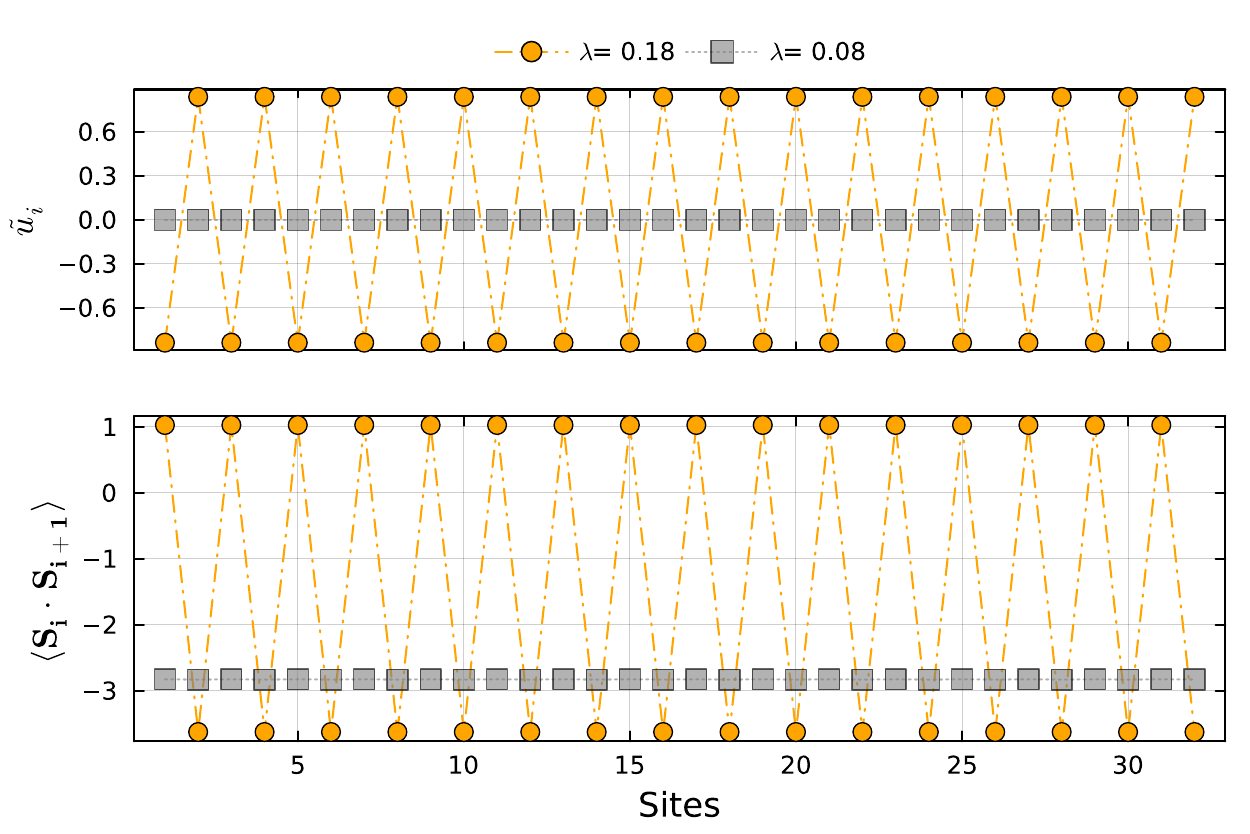}
    \caption{Elastic displacements (top panel) and spin correlations (bottom panel) for two representative values of the spin-phonon coupling $\lambda$. 
    Gray squares: $\lambda=0.08 < \lambda_c$.
    Orange circles: $\lambda=0.18 >  \lambda_c$.}
    \label{fig:sample-deltas-and-correlations}
\end{figure}

In the weak coupling regime (lowest $\lambda$ in the figure) it is numerically difficult to distinguish whether the minimum energy is obtained for a homogeneous configuration or 
for a slight dimerization with $\tilde{u}_0 \ll 1$ 
(see in Fig.  \ref{fig:sample-deltas-and-correlations}, top panel, 
the presumably absence of distortions for $\lambda=0.08$ and discussion in Section IV).
In contrast, for strong coupling (largest $\lambda$ in Fig.\ \ref{fig:energy-landscapes})   
a highly distorted phase shows up with $\tilde{u}_0$ of the order of 0.5 
(see details in Fig.\  \ref{fig:sample-deltas-and-correlations}, top panel, for  $\lambda=0.18$).
In the latter case physically meaningful distortions ${u}_i = \tilde{u}_i/ \alpha$, 
which should be much smaller than  the lattice spacing $a$,  require materials with $\alpha \gg 1/a$.

We describe below that there is a critical value of $\lambda$ where a 
first-order transition takes place.
The transition to the strongly dimerized behavior 
corresponds to a second local energy minimum becoming more favorable than the expected homogeneous or weakly-dimerized configuration,
as illustrated in Fig.\ \ref{fig:energy-landscapes} by results at, below and above the critical value, 
estimated as $\lambda_c \approx 0.1355$. 
In the figure, we plot nearly coinciding energies for $N=20$ and 50 sites. The results were checked to remain  stable in chains of $N=10, \,30$ and $40$ sites as well.
One can see the presence of two local minima, one for vanishing or tiny $\tilde{u}_0$ and another for large $\tilde{u}_0$.
The minimum for large displacements becomes energetically favorable at strong coupling 
$\lambda$.
We have thus found a quantum first-order transition as a function of the spin-phonon parameter 
$\lambda$, a phenomenon not present in the spin $S=1/2$ magnetoelastic chain.

\begin{figure}
    \centering
    \includegraphics[width=1.0\linewidth]{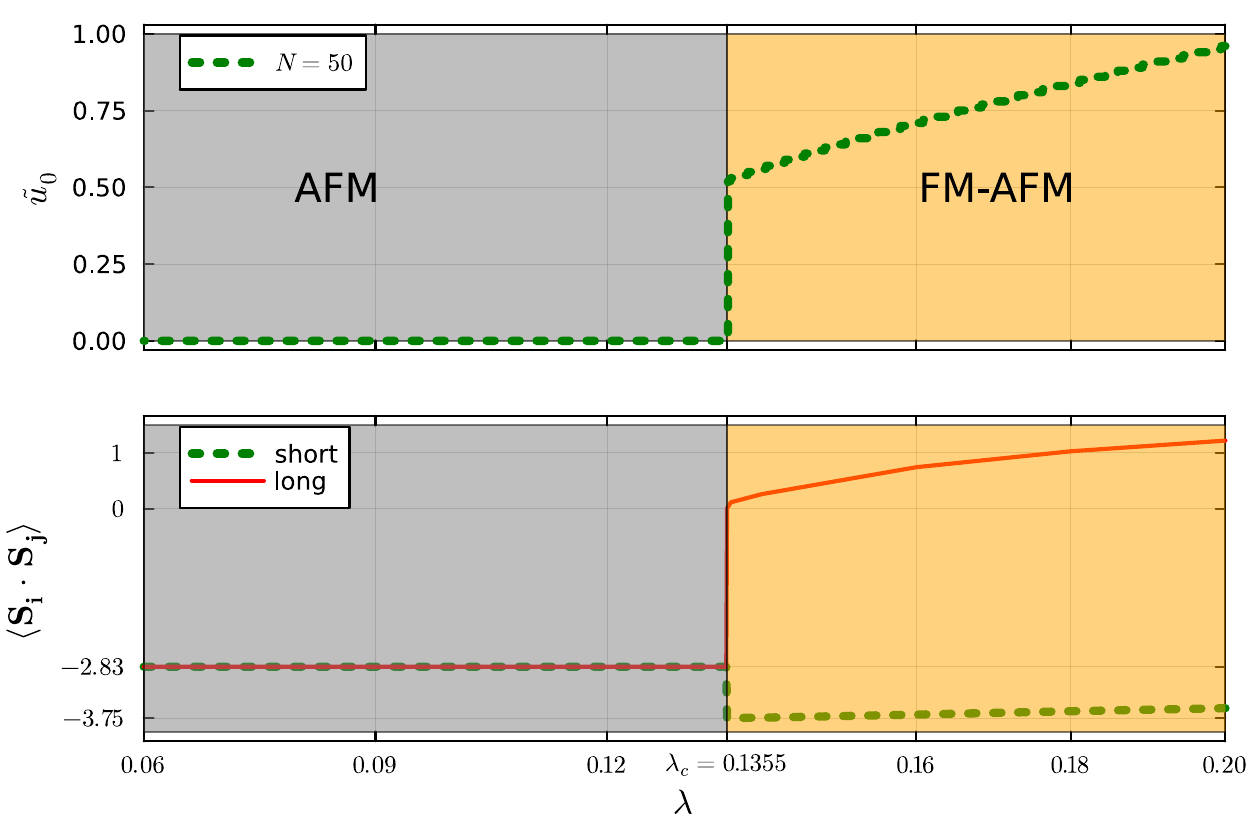}
    \caption{Amplitude of alternating displacements (top) and spin-spin correlations (bottom) in the ground-state  as a function of the spin-phonon coupling.}
    \label{fig:u_0-vs-lambda}
\end{figure}

 In Fig.\ \ref{fig:u_0-vs-lambda}, top panel, we plot the amplitude of alternating displacements versus $\lambda$
 in the ground state of the system, for a sample chain length. 
 One can clearly see a jump, occurring at 
 $\lambda_c \approx 0.1355$. 
 Within our numerical precision the value of 
 $\lambda_c$ is not sensitive to the chain length, 
 a robust feature that should be valid in the thermodynamic limit.  
 Beyond the critical point the displacement amplitudes jump to values larger than 0.5, 
 making the exchange $J_i = J \left[ 1-(\tilde{u}_{i+1}-\tilde{u}_{i} )  \right]=J ( 1-2\tilde{u}_0)$  
 small and negative at bonds where ions become separated. 
 This generates an important alternation of the spin exchange between 
 strong antiferromagnetic dimers (at short bonds) and weaker ferromagnetic interactions (at long bonds), 
 producing a deep effect in the correlations in the strong coupling phase. 
 In the bottom panel of Fig.\ \ref{fig:sample-deltas-and-correlations} we show the nearest neighbors spin-spin correlations. 
 They alternate between  short and long bonds:
 at each short bond the correlation indeed takes a value close to 
 $-\frac{3}{2}\left(\frac{3}{2}+1\right)=-\frac{15}{4}$, which is the value corresponding to perfect singlet dimer states. 
 For the long bonds the spin-spin correlations are in general positive (ferromagnetic).
 Importantly, at the transition we find $\tilde{u}_0=0.5$, so that perfect singlet dimers form in the AFM bonds with correlations  $-\frac{15}{4}$, 
 exactly decoupled from each other.
 
Were we analyzing a different magneto-elastic model, usual elsewhere [\onlinecite{Dobry-Riera_1995,Feiguin_1997,Penc_2004,BPM-un-medio,Poilblanc}], 
where the elastic energy is associated to bond distortions with an ad-hoc fixed length constraint (the so called bond-phonon model),
we would find a spurious phase separation [\onlinecite{disclaimer}]. Such non-realistic behavior has not been observed for spin $S=1/2$ systems. 
The ESP studied here is more realistic in automatically taking into account the unavoidable correlations between bond lengths arising
from the underlying three-dimensional lattice environment (see for instance [\onlinecite{Vishwanath_2007}]).

\subsubsection{Magneto-elastic response to external magnetic fields}

\paragraph{Spin gap:}

As is well known the dimerization of exchange couplings produces a finite spin gap,
defined in the present model as the difference of the total (magnetic plus elastic) energy 
between the lowest lying states with $S^z_\text{total}=1$ and $S^z_\text{total}=0$,   denoted here as $\Delta_{tot}$.

Given the abrupt onset of alternating distortions described in the previous Section, 
the spin gap is expected to jump from being zero or very small (see Section \ref{low-lambda}) 
to a finite value in the FM-AFM phase (see Section \ref{sect:perturbations}). 
 %
 \begin{figure} %
 	\centering
 	\includegraphics[width=1.0\linewidth]{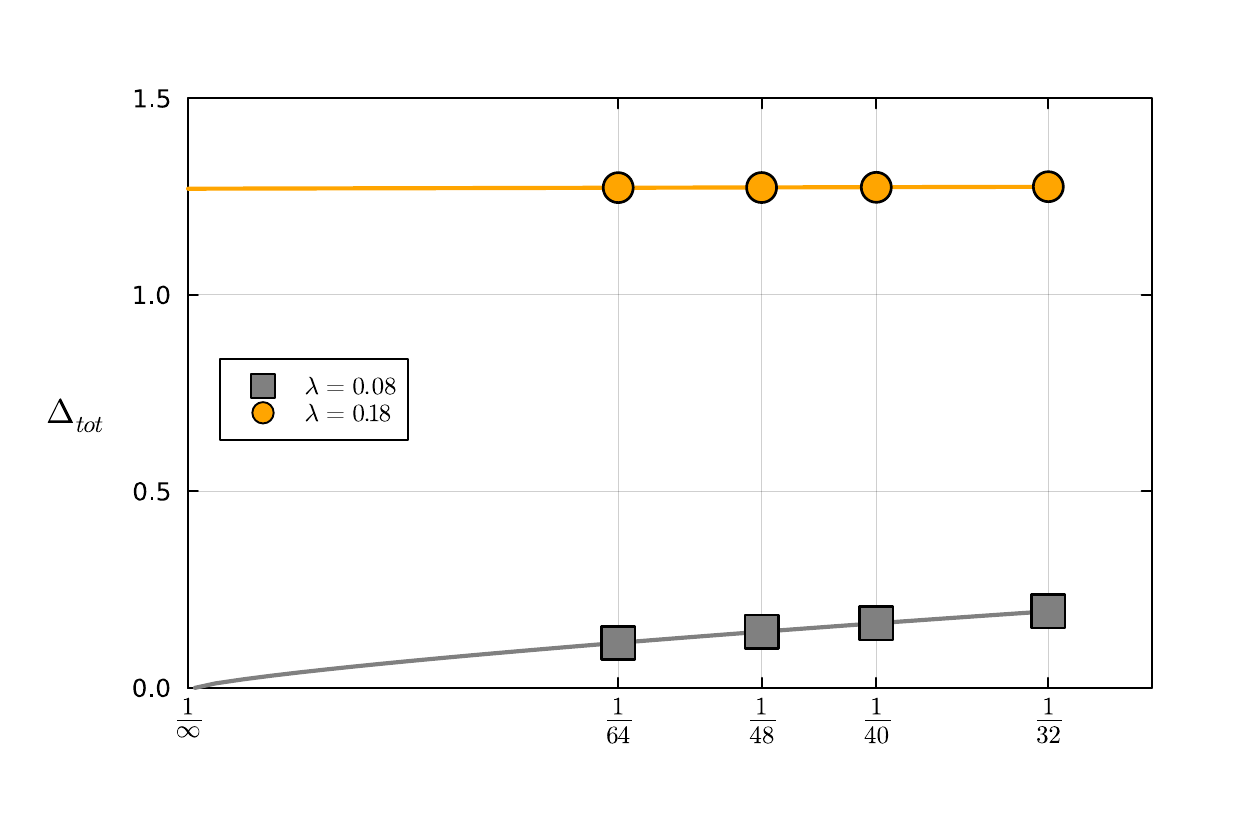}
 	\caption{Finite size scaling of the spin gap for two values  of $\lambda$ at opposite sides of the transition 
 	$\lambda=0.08$ (circles) and $\lambda=0.18$ (squares).
 }
 	\label{fig:total-spin-gap}
 \end{figure}
 %
 We show in Fig.\ \ref{fig:total-spin-gap} a finite size scaling of the spin gap $\Delta_{tot}$ 
 for the representative values of the spin-phonon coupling in each phase selected in previous figures,
 $\lambda=0.08<\lambda_c$ and $\lambda=0.18>\lambda_c$. 
 Solid curves correspond to $1/N^\nu$ fitting.
 Below $\lambda_c$ the extrapolated gap is zero within numerical resolution, while it is finite and almost size independent in the FM-AFM phase.

\paragraph{Magnons, solitons and magnetization plateaus in the highly distorted phase:}

In the highly distorted phase ($\lambda>\lambda_c$) the state with $S^z_\text{total}=1$ exhibits dimerized distortions and spin correlations, 
very similar to the ground state $S^z_\text{total}=0$, with  a homogeneous distribution of the magnetization along the chain. 
The same happens in the sector $S^z_\text{total}=2$. 
That is, the magnetic excitations behave as delocalized magnons.
 
 \begin{figure} %
 	\centering
 	\includegraphics[width=1.0\linewidth]{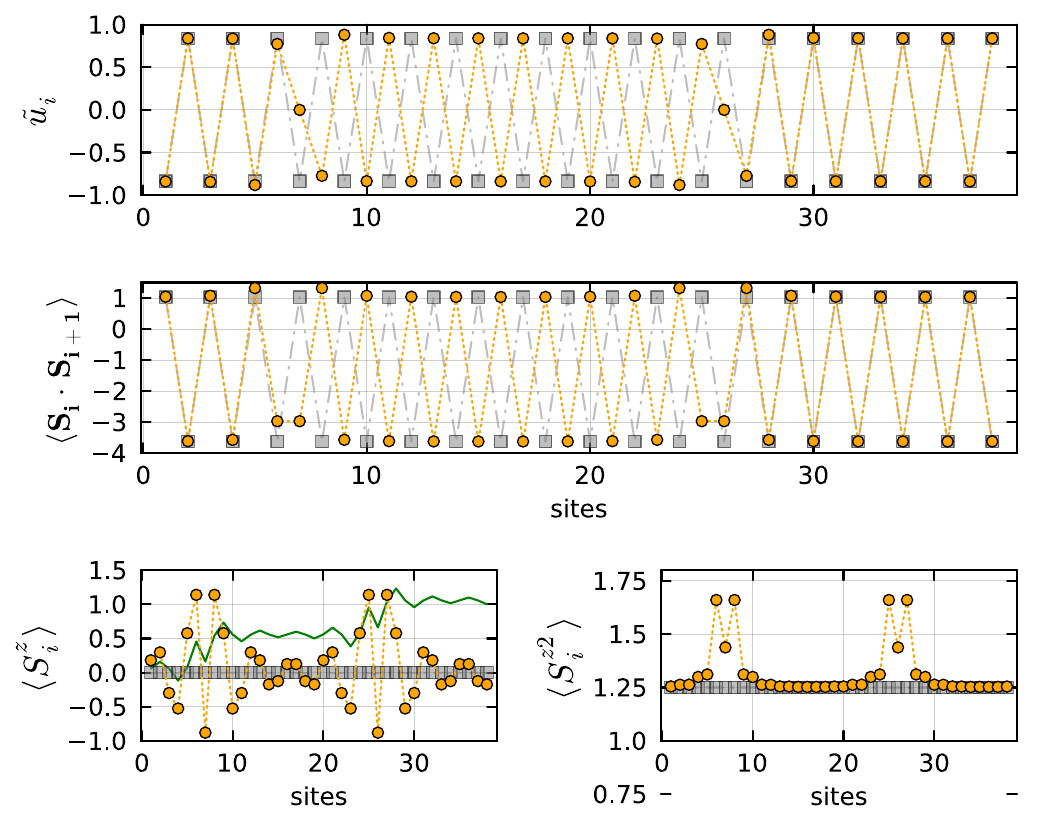}
 	\caption{Comparison of observables between the lowest energy states with $S^z_\text{total}=3$ (orange circles) and $S^z_\text{tot}=0$ (gray squares), for $\lambda=0.18>\lambda_c$. Top panel: elastic ion displacements. Middle panel: nearest neighbors spin-spin correlations. Left bottom: local expectation value of the spin projection; the green line describes the accumulated magnetization (normalized to one). Right bottom: local spin anisotropy.}
 	\label{fig:S_z=3}
 \end{figure}
 
In contrast, we have found that reaching $S^z_\text{total}=3$ (that would correspond to a rigid flip of a $S=3/2$ classical vector), 
domain walls (solitons) appear separating regions with elastic patterns that differ in that the displacements are shifted by one site.
This is shown  in Fig.\ \ref{fig:S_z=3}, where we compare observables in the $S^z_\text{total}=3$ (orange circles) and $S^z_\text{total}=0$ (gray squares) ground states. 
In the top panel we show how ion displacements reconfigure with respect to the $S^z_\text{total}=0$ profile
taking in different regions the two possible (one-site shifted) alternating patterns. These dimerized domains are separated by two sharp domain walls.
This behavior resembles the splitting of $S^z_\text{total}=1$ magnon into a pair of $S^z=1/2$ localized excitations (spinons) taking place in the 
spin $S=1/2$ magneto-elastic chain [\onlinecite{Poilblanc}]. There is however
a difference in the detailed shape of these  soliton discussed below.

In the middle panel we show the corresponding spin-spin correlations,
which follow the alternation of short (antiferromagnetic) and long (ferromagnetic) bonds and the width of the domain walls. 

In the left bottom panel we report the expectation value of the local $S^z_i$ spin component. 
One can see that the contributions to $S^z_\text{total}=3$
are localized around the domain walls, with a  characteristic soliton shape [\onlinecite{Temo}]. 
The green line, indicating the accumulated $\langle S^z \rangle$ contribution (normalized to one), 
shows that each soliton carries $S^z=3/2$. 
Moreover, in the right bottom panel the local expectation value of $(S^z)^2$ 
($1.25$ for each component in the $S^z_\text{total}=0$ isotropic case) shows a spontaneous single-ion anisotropy, 
towards the $z$ component in each soliton. 
These solitonic features in the $S=3/2$ spin (so similar to the $S=1/2$ case) 
suggest the existence of a $S^z=3/2$ spinon and deserve further investigation to be published elsewhere [\onlinecite{elsewhere}].

From the displacements in Fig.\ \ref{fig:S_z=3}, top panel, 
one can see that the bonds around the center of domain walls are \textit{short}, 
leading to three closer sites separating two nearly perfectly dimerized regions. We call this case B. 
On the other hand the simplest defect that could separate two dimerized regions, say case A, 
would be an isolated spin 3/2 separated by  \textit{long} bonds to the neighboring dimerized regions. 
While the elastic cost looks the same, 
case A gains magnetic energy in one more dimerized bond (AFM singlet) with respect to case B, 
but case B can gain energy by arranging the spins of the three sites. 
To estimate the energy balance 
assume that the displacements are $\tilde{u}=\pm 0.5$ (near the transition) so that $J_i=0$ ($2J$) for the long 
(short) bonds in the dimerized phases. 
To first order the singlets do not interact with the domain wall structure in the middle. 
In case B one has a chain of three
sites interacting with $J_B=3/2\,J$. The energy is minimized coupling ferromagnetically the two spins at the borders leading to a total spin $S_b=3$ 
and coupling the resulting spin antiferromagnetically with the central one $S_c=3/2$,
leading to a total spin $S_t=3/2$. 
A simple calculation leads to $\mathbf{S_b} \cdot \mathbf{S_c} = - S_b(S_b+1)/2=-6$ and a gain of energy $E_3=-9J$ 
which overcomes the loss of one singlet energy   $15/2\, J$, so that case B is favored. 
Furthermore, since the spin of 
any of the border spins is $\mathbf{S_b}/2$ their correlation with the central one is -3 
in perfect agreement with Fig.\ \ref{fig:S_z=3}, middle panel. 
Full diagonalization of the Hamiltonian for three isolated sites 
shows that the  ground state indeed belongs to the irreducible representation with $S_t=3/2$
and that the local expectation values  $\langle S^z_i \rangle$ and $\langle (S^z_i)^2 \rangle$ are very close 
to those shown in Fig.\ \ref{fig:S_z=3}, bottom panels;
this also explains that the domain wall structure first shows up in the $S^z_\text{total}=3$ sector.
An analogous calculation for the spin-1/2 chain leads to $E_3=-3/2 \, J$ which
is smaller than the energy loss of one singlet ($3J$), and therefore case A (the well known $S^z=1/2$ spinons)
is favored.

\begin{figure} %
	\centering
	\includegraphics[width=1.0\linewidth]{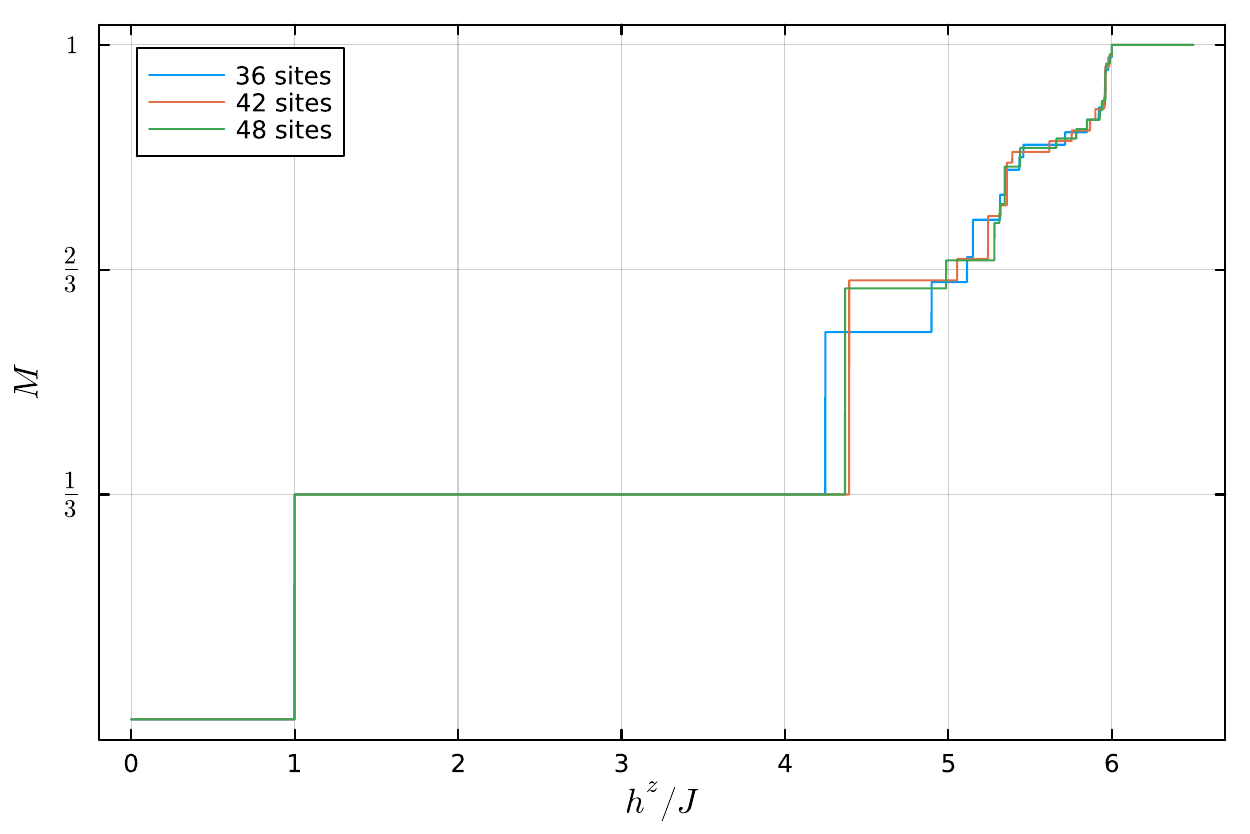}
	\caption{Magnetization curves of the magneto-elastic model, for $\lambda=0.18>\lambda_c$ 
	and three different chain lengths. 
	Saturation magnetization is normalized to one. }
	\label{fig:magnetization curve}
\end{figure}

\begin{figure} %
	\centering
	\includegraphics[width=1.0\linewidth]{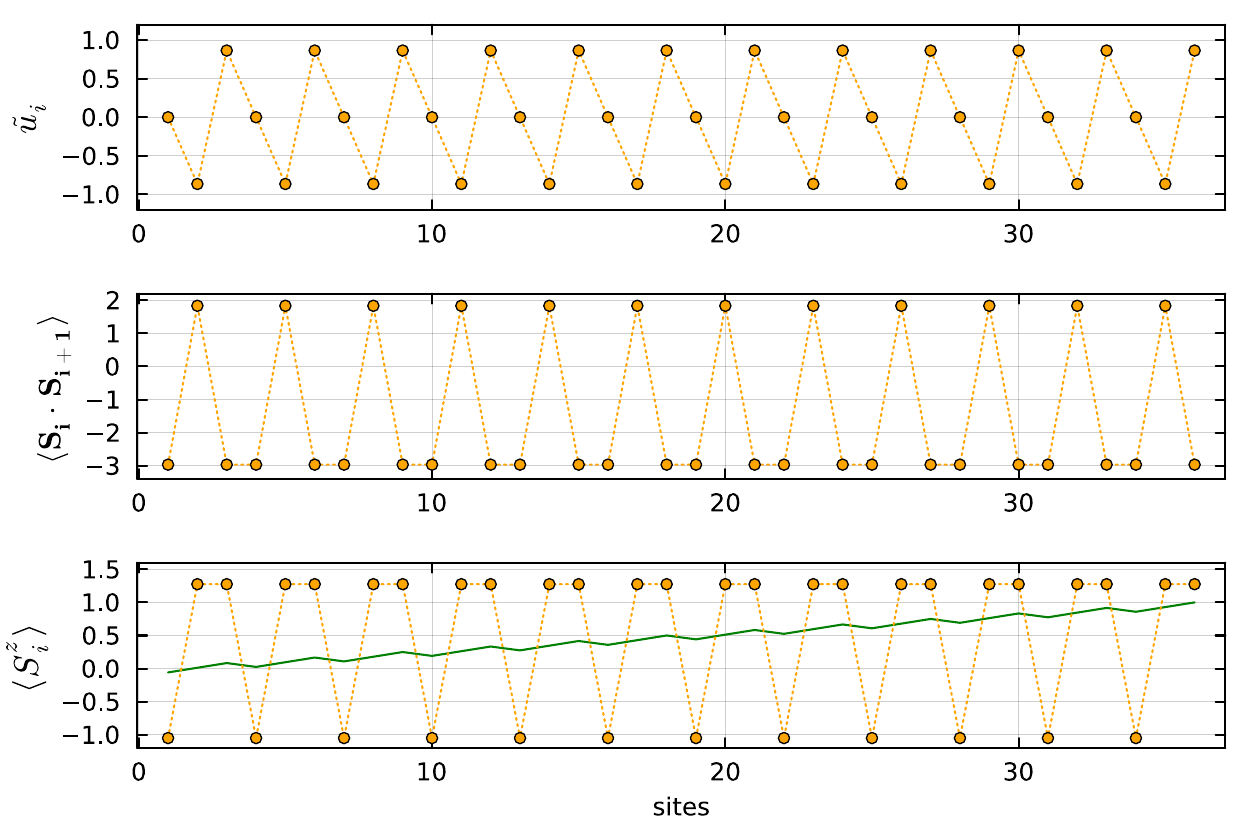}
	\caption{Local ion displacements, spin-spin correlations and local magnetization for the $M=1/3$ state, 
	for $\lambda=0.18>\lambda_c$. 
	One can see that the domain walls present in Fig.\ \ref{fig:S_z=3} now fill up the whole chain.
	We show in detail the $S^z_\text{total}=18$ lowest energy state in a chain of 36 sites.}
	\label{fig:M=1/3}
\end{figure}

Increasing further the magnetic field, 
the magnetization curve in Fig.\ \ref{fig:magnetization curve} shows another interesting feature.
One finds a wide zero magnetization plateau that is stable until a metamagnetic transition takes place. 
This means that, upon increasing the external magnetic field $h^z$, 
at some point the Zeeman energy makes favorable some high magnetization state 
(with not too high energy) before selecting either the $S^z_\text{total}=1$ or the 
$S^z_\text{total}=3$ state described above.
We have found that the magnetization jumps  directly to  a finite magnetization fraction $M=1/3$ 
(where $M=1$ means saturation). 
In this state the discussed $S^z=3/2$ spinons condensate every three sites occupying the length of the whole chain. The corresponding observable profiles are shown in Fig.\ \ref{fig:M=1/3}.

The presence of a plateau and a metamagnetic jump is similar to the one found in the $S=1/2$ magneto-elastic chain [\onlinecite{Temo,Poilblanc}], where bosonization arguments do explain the behavior.
In contrast to that case, now the spin gap and the metamagnetic jump are seen only when $\lambda > \lambda_c$ which is far from the perturbative region where bosonization could be applied.

The $M=1/3$ ordered structure is robust against the magnetic field and gives rise to a second, wider, $M=1/3$ plateau.  
For larger fields, it is not clear to us whether the magnetization curve develops smoothly or 
shows a third $M=2/3$ plateau
(further investigation to be published elsewhere [\onlinecite{elsewhere}]).

\section{Analytical insight from perturbation theory}
\label{sect:perturbations}

\subsection{Perturbation theory in the exchange at the weak links}
\label{subsect:perturbative-energy}

Motivated by the almost decoupled dimer structure found and described in the previous Section, we consider a periodic Heisenberg chain for spin $S=3/2$ with alternating exchange
interactions. 
\begin{eqnarray}
H &=&H_{1}+H_{2},  \nonumber \\
H_{1} &=&J_{1}\sum\limits_{j=0}^{N/2-1}\mathbf{S}_{2j+1}\cdot \mathbf{S}_{2j+2},  \nonumber \\
H_{2} &=&J_{2}\sum\limits_{j=0}^{N/2-1}\mathbf{S}_{2j+2}\cdot \mathbf{S}_{2j+3}.  \label{htot}
\end{eqnarray}

For the alternating ion displacement found numerically, the exchanges are
$J_{1}=J (1+2\tilde{u}_0  )$, $J_{2}=J (1-2\tilde{u}_0 )$. Then the values $\tilde{u}_0$ close to 0.5 found in the strongly dimerized phase correspond to the perturbative regime $|J_{2}|\ll J_{1}$. 
%
\begin{figure} 
    \centering
    \includegraphics[width=0.9\linewidth]{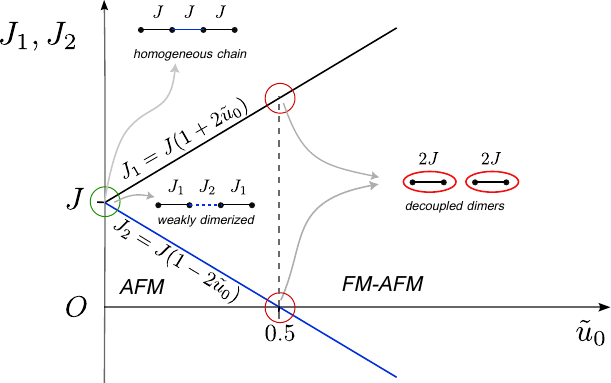}
    \caption{Dimerized exchange couplings $J_1,\,J_2$ in terms of the displacement amplitude $\tilde{u}_0$. 
    The null/tiny distortions found for $\lambda < \lambda_c$ (see Fig.\ \ref{fig:u_0-vs-lambda}) 
    lead to a homogeneous/weakly-dimerized antiferromagnetic spin chain (green circle region), 
    while $\tilde{u}_0>0.5$ found for $\lambda>\lambda_c$ (red circles region) 
    leads to an  alternating (strong) antiferro- (weak) ferromagnetic spin Hamiltonian. 
    In this phase, close to the transition, $|J_2|/J_1 \ll 1$ allows for perturbative analysis 
    around the decoupled dimers product state (dashed line region). }
    \label{fig:J1-J2}
\end{figure}

The ground state to be perturbed, 
for $J_{2}=0$ ($\tilde{u}_0 \equiv 0.5$),  
consists of a direct product of singlets located on 
short bonds 1-2, 3-4, 5-6 ... 
To be precise, it is useful to use the index $j$ in Eq.\ (\ref{htot}) to label the $J_1$  bonds, 
denoting  by $|D_j:S,m\rangle $, the eigenstates of $J_{1}\mathbf{S}_{2j+1}\cdot \mathbf{S}_{2j+2}$ with total spin $S$ and spin projection $m$. 
The dimer singlet on bond $j$,  $|D_j:0,0 \rangle $, can be
straightforwardly constructed using the Clebsch-Gordan coefficients that
combine two spins 3/2 to build a total spin $S=0$. Similarly,
excited states with total spin $S=1,\,2,\,3$ can be constructed for each bond $j$. 
The energy of these states is given by 
$E_{S}=J_{1} [ S(S+1)/2-15/4]$. 
Finally the ground state of the whole chain, for $J_{2}=0$, reads
\begin{equation}
| g \rangle =\bigotimes_{j=0}^{N/2-1}  |D_j:0,0 \rangle .
\label{decoupled gs}
\end{equation}

One can easily see that the only non-vanishing terms to be evaluated in a perturbative computation of the ground-state energy are those involving matrix elements of the individual terms in $H_2$, proportional to $J_{2}$. 
Let us discuss the action of one of them,
for example  $H_{23}=J_{2}\,\mathbf{S}_{2}\cdot \mathbf{S}_{3}$ on the ground state, which only acts on the neighboring dimers $|D_0:0,0\rangle$ and $|D_1:0,0\rangle$.  After a lengthy
but straightforward algebra we obtain the simple result

\begin{eqnarray}
&&H_{23}|D_0:0,0\rangle |D_1:0,0\rangle  =\frac{5\sqrt{3}J_{2}}{4}|e_{0,1}\rangle ,
\nonumber \\
&&|e_{0,1}\rangle  =\frac{1}{\sqrt{3}} ( |D_0:1,1\rangle |D_1:1,-1\rangle -|D_0:1,0\rangle|D_1:1,0\rangle 
\nonumber \\
&&+|D_0:1,-1\rangle |D_1:1,1\rangle ) .  
\label{me}
\end{eqnarray}
The form of the result is expected. The state $|e_{0,1}\rangle$, that is the excitation between dimers 0 and 1,  is a total
singlet in sites 1, 2, 3, 4 obtained combining two triplets $S=1$ in sites 1, 2  and sites 3, 4. The vector operator  $\mathbf{S}_{j}
$ has matrix elements only between states that differ in $S$ by at most 1
(exactly 1 when applied to the singlet). On the other hand,  since $H_{23}$
is invariant under rotations, the singlet character of the ground state is
conserved after its application to the ground state. 

\begin{figure}[hb]
\centering
\includegraphics*[width=1.0\columnwidth]{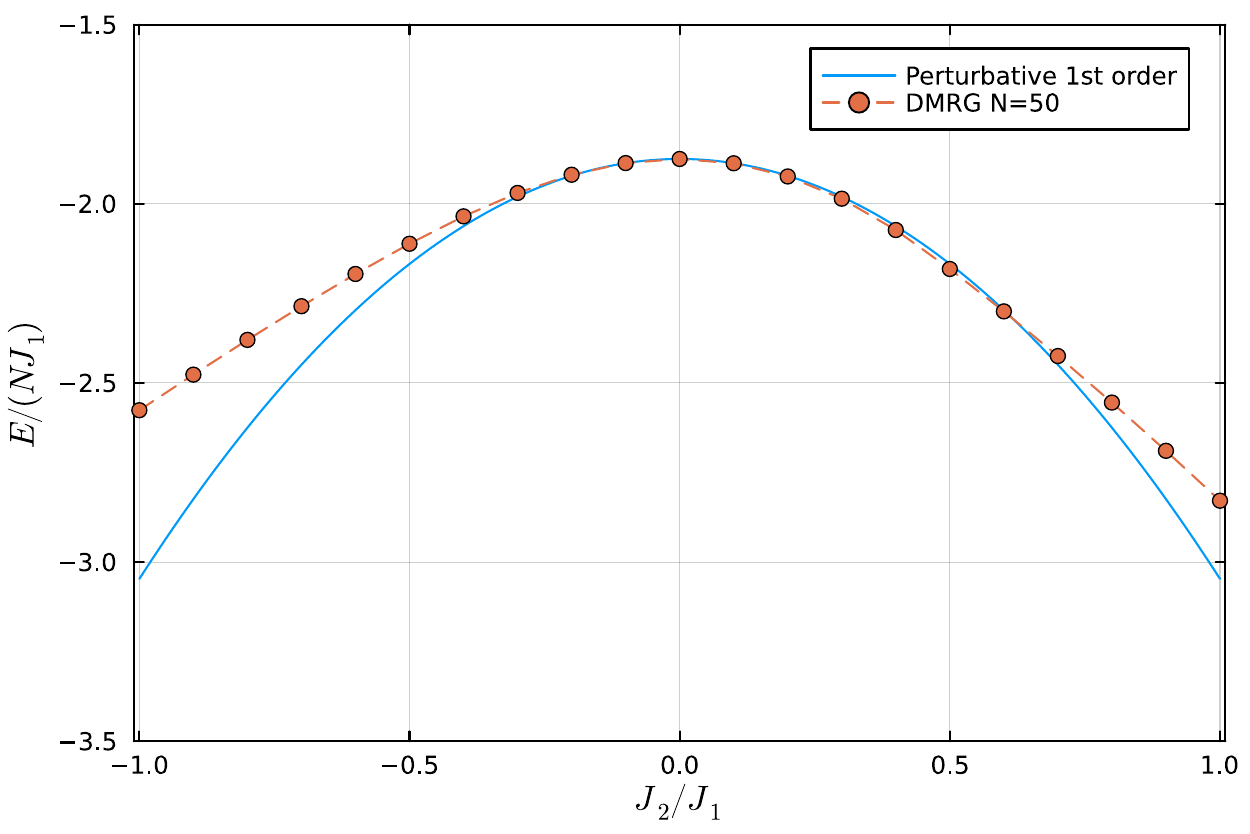}
\caption{Comparison of DMRG results for the ground-state energy per site in 
a ring of 50 sites and the corresponding perturbative ones.}
\label{eneper}
\end{figure}

Using the standard equation for the energy in second order perturbation
theory 

\begin{equation}
E=E_{g}-\sum\limits_{e}\frac{|\langle e|H_{2}|g\rangle |^{2}}{E_{e}-E_{g}},
\label{pert}
\end{equation}
where $|g\rangle $ is the ground state in Eq. (\ref{decoupled gs}) and $E_{g}$ its energy,
and similarly $|e\rangle $, $E_{e}$ for the excited states, the energy per
site becomes

\begin{equation}
E_{\text{spin}}=-\frac{15}{8}\left( J_{1}+\frac{5J_{2}^{2}}{J_{1}}\right) .
\label{ene}
\end{equation}
While this result is quantitatively valid only for $|J_{2}|\ll J_{1}$, it
remains qualitatively valid over a wider range, as illustrated in Fig. \ref{eneper}.

Returning to the dimerized magneto elastic chain, adding the
elastic energy $\tilde{u}_0  ^{2}/(2\lambda )$, the total energy has the form
\begin{equation}
E_{\text{total}}/J=-A\left[ 1+2\tilde{u}_0  +B\frac{(1-2\tilde{u}_0  )^{2}}{1+2\tilde{u}_0  }\right]
+\frac{\tilde{u}_0^{2}}{2\lambda },  \label{etotal}
\end{equation}
where $A=15/8$ and $B=5/8$. Expanding in series of $\tilde{u}_0  $ one
obtains that $E_{\text{total}}$ can also be written in the form 
\begin{eqnarray}
E_{\text{total}}/J &=&-A(1+B)+2A(3B-1)\tilde{u}_0   \nonumber \\
& &  +8(\frac{1}{\lambda }-2AB)\tilde{u}_0^{2} + 32AB\frac{\tilde{u}_0 ^{3}}{1+2\tilde{u}_0 }.  \label{eexp}
\end{eqnarray}
This expression has a positive linear term, indicating that a small
dimerization is not expected in this chain. But for large enough 
$\lambda $, the quadratic term gets negative and a first-order transition is
expected, in agreement with the numerical results.
The last positive term of Eq. (\ref{eneper}) ensures the stability of the system for large $\tilde{u}_0 $.
%

We have also performed the perturbative calculations for the chain with spin
1/2. The algebra is much simpler and the results are similar. In this case, $A=3/8$ and $B=1/4$. As a consequence of the smaller value of $B$, the
term linear in $\tilde{u}_0 $ becomes negative. This results in a finite
dimerization for any $\lambda$, hence no first-order transition takes place, 
as it is well known [\onlinecite{Affleck_1989}].

\subsection{Correlation functions}
\label{subsect:correlations}

The spin-spin correlation function for
the chain with spin 3/2, computed as first order, provide just linear corrections.
However,
the matrix
elements of Eq. (\ref{me}) for the interactions in the weak links with
exchange $J_{2}$ are very large, then quadratic terms are important. 
To consider them in a simple analytical way, 
we have performed variational calculations in
small clusters, which include the linear terms in $J_{2}$ calculated with 
Eq. (\ref{me}), but incorporate second-order terms, in particular taking
into account the norm of the variational state.

To calculate the spin correlation in a weak link, say 
$\left\langle \mathbf{S}_{2i+2}\cdot \mathbf{S}_{2i+3}\right\rangle =\left\langle 
\mathbf{S}
_{2}\cdot \mathbf{S}_{3}\right\rangle $, we consider a system of four sites
1 to 4, in which the spins 1-2 and 3-4  interact via a $J_{1}\mathbf{S}_{1}\cdot 
\mathbf{S}_{2}$ and $J_{1}\mathbf{S}_{3}\cdot \mathbf{S}_{4}$, while 2-3
have an interaction $J_{2}\mathbf{S}_{3}\cdot \mathbf{S}_{4}$. We restrict
to the Hilbert space of two states: the first one, which we shortly denote as 
$|1\rangle =|S_{12}\rangle |S_{34}\rangle \equiv $ \break $ |D_0:0,0\rangle |D_1:0,0\rangle $, is
the product of the singlets 1-2 and 3-4 described above Eq. (\ref{me}). The
second one denoted by $|2\rangle =|S_{1-4}\rangle =|e\rangle $ is the
singlet obtained combining triplets 1-2 and 3-4 written in Eq. (\ref{me}).
Using Eq. (\ref{me}), and 

\begin{equation}
\alpha =\left\langle 2|\mathbf{S}_{2}\cdot \mathbf{S}_{3}|2\right\rangle
=-2(2-\sqrt{3})/5,  \label{alp}
\end{equation}
obtained after some algebra, the Hamiltonian restricted to this subspace 
takes the form of the following matrix
\begin{equation}
\left( 
\begin{array}{cc}
-\frac{15}{2}J_{1} & \frac{5\sqrt{3}}{4}J_{2} \\ 
\frac{5\sqrt{3}}{4}J_{2} & -\frac{11}{2}J_{1}+\alpha J_{2}%
\end{array}
\right) .  \label{maw}
\end{equation}
The ground state is $|g\rangle =u|1\rangle -v|2\rangle $, with 
$u,v>0$, $u^{2}+v^{2}=1$  and 

\begin{equation}
v^{2}=\frac{1}{2}-\frac{J_{1}+\alpha
J_{2}/2}{2\sqrt{(J_{1}+\alpha
J_{2}/2)^{2}+75J_{2}^{2}/16}}.  \label{vc}
\end{equation}
Using again  Eqs. (\ref{me}) and (\ref{alp}), the correlation in the weak
links becomes 

\begin{equation}
\left\langle \mathbf{S}_{2}\cdot \mathbf{S}_{3}\right\rangle =-\frac{5\sqrt{3}}{2}uv+\alpha v^{2}.  \label{s23}
\end{equation}

\begin{figure}[h]
\begin{center}
\includegraphics[width=1.0\columnwidth]{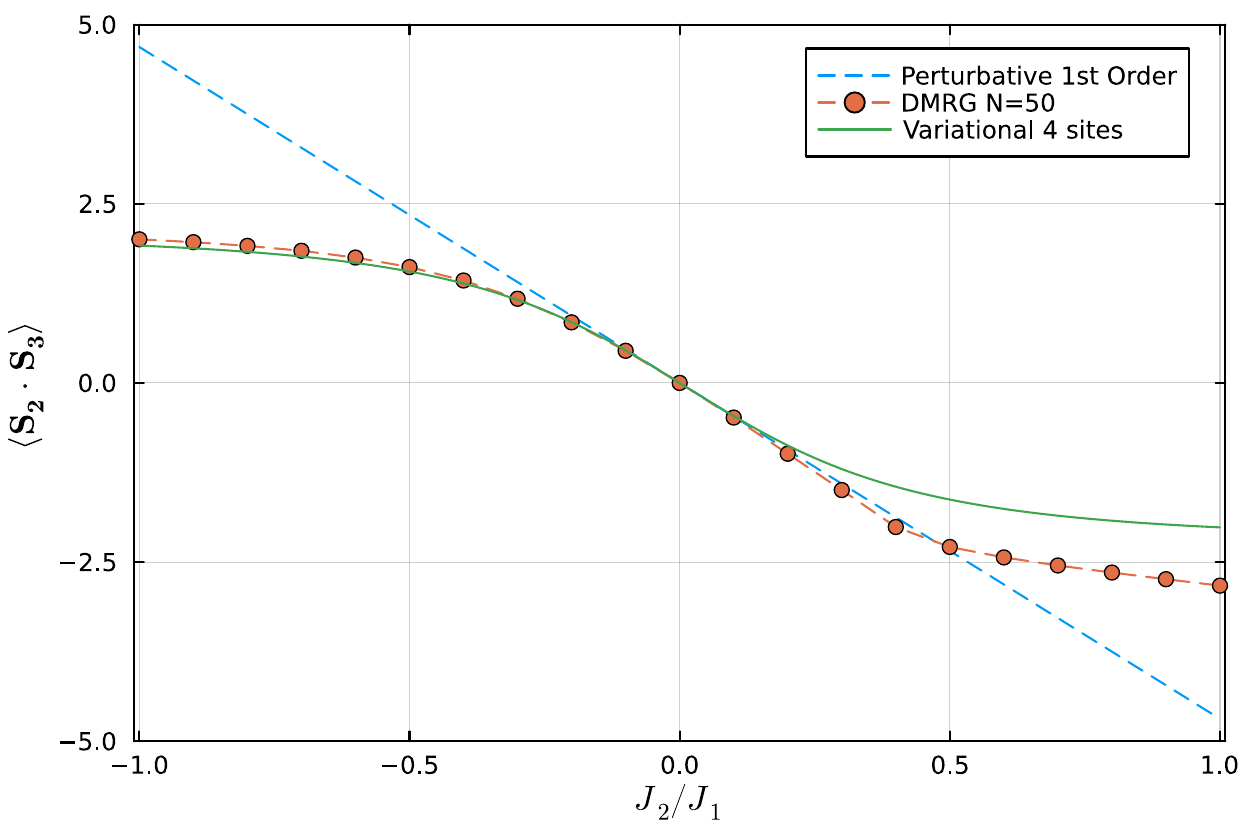}
\end{center}
\caption{Spin-spin correlation
function in the weak links as a function of $J_2/J_1$.}
\label{corrw}
\end{figure}

The result given in Eq. (\ref{s23}) is represented in Fig.
\ref{corrw} and compared with the first-order
perturbative result $-75 J_2/(16 J_1)$ and DMRG data. The improvement
with respect to first-order perturbation is noticeable.
In particular, the agreement for negative $J_2$ 
with the DMRG result is remarkable.

To obtain an approximation for the correlation of the strong links  
one needs to
follow a similar procedure as above, but for a chain of 6 sites to calculate 
$\left\langle \mathbf{S}_{3}\cdot \mathbf{S}_{4}\right\rangle $. Now the two
relevant states have the form 

\begin{eqnarray}
|1\rangle  &=&|S_{12}\rangle |S_{34}\rangle |S_{56}\rangle ,  \label{statess}
\\
|2\rangle  &=&\frac{1}{\sqrt{2}}\left( |S_{1-4}\rangle |S_{56}\rangle
+|S_{12}\rangle |S_{3-6}\rangle \right) ,  \nonumber
\end{eqnarray}
leading to the matrix 

\begin{equation}
\left( 
\begin{array}{cc}
-\frac{45}{4}J_{1} & \frac{5\sqrt{6}}{4}J_{2} \\ 
\frac{5\sqrt{6}}{4}J_{2} & -\frac{37}{2}J_{1}
+\alpha J_{2}
\end{array}
\right) .  \label{mas}
\end{equation}
Now the coefficient of the ground state is modified to

\begin{equation}
v^{2}=\frac{1}{2}-\frac{J_{1}+\alpha
J_{2}/2}{2\sqrt{(J_{1}+\alpha
J_{2}/2)^{2}+150J_{2}^{2}/16}},  \label{vcs}
\end{equation}
and 

\begin{equation}
\left\langle \mathbf{S}_{3}\cdot \mathbf{S}_{4}\right\rangle =-\frac{15}{4}u^{2}-\frac{11}{4}v^{2}.  \label{s34}
\end{equation}

\begin{figure}[h]
\begin{center}
\includegraphics[width=1.0\columnwidth]{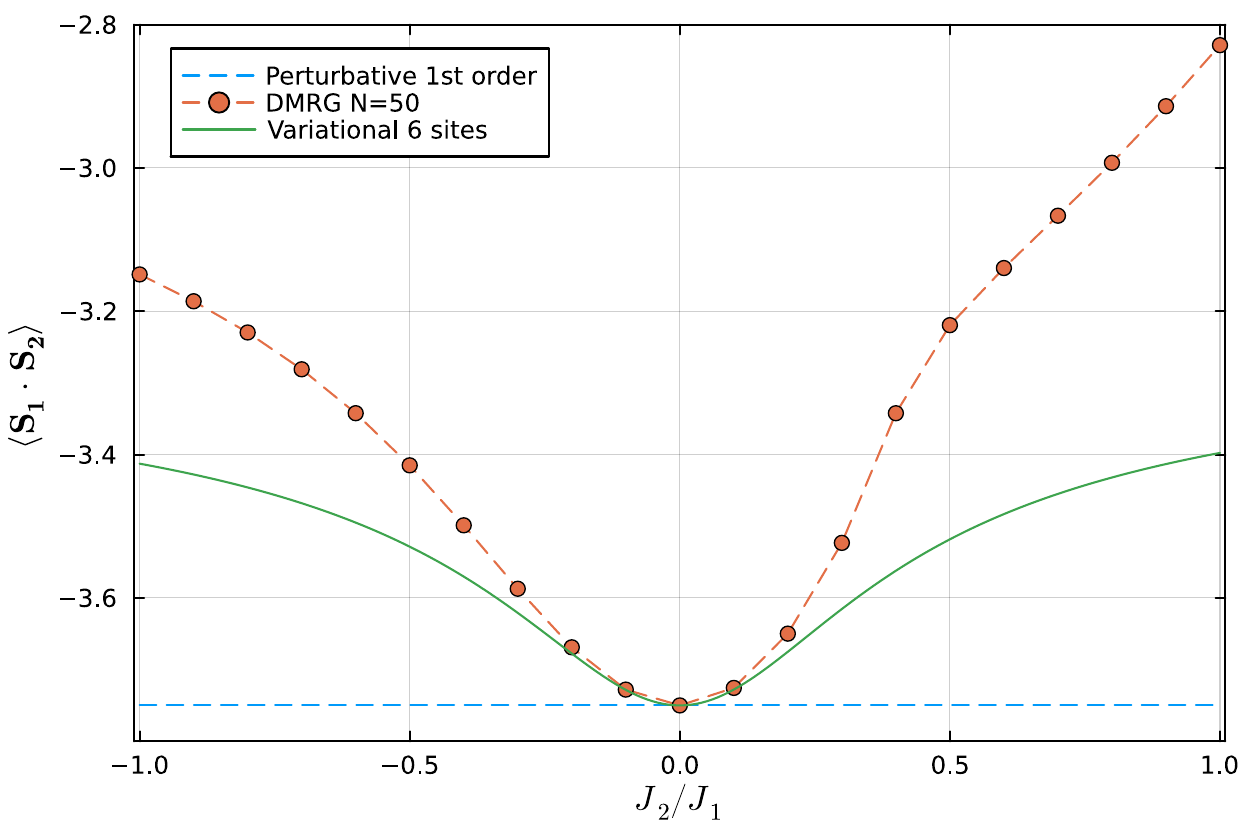}
\end{center}
\caption{Spin-spin correlation
function in the strong links as a function of $J_2/J_1$.}
\label{corrs}
\end{figure}

The result of Eq. (\ref{s34}) is represented in Fig.\
\ref{corrs} and compared with DMRG calculations. 
To linear order in perturbation theory, there is no change in the correlation function with respect to the case $J_2=0$. The improvement introduced by the variational calculation is only qualitative in this case, 
indicating that it would be necessary to incorporate other second-order corrections to improve the approximation.

\subsection{Spin gap}
\label{subsect:gap}

To compute the energy gap, we calculate the effect of $J_{2}$ up to second
order in the energy of the triplet state in which all of the $J_{1}$ bonds are in the singlet ground state, except one (denoted as $B$), which
is in the triplet state $|B:1,1\rangle $. Due to the $SU(2)$ symmetry of the
model, the energy is the same for the states with other spin projections $|B:1,m\rangle $. After another lengthy calculation, a similar procedure that
led to equation (\ref{me}) gives

\begin{eqnarray}
&&(H_{23}/J_{2})|A:0,0\rangle |B:1,1\rangle =-\frac{5}{4}|\alpha 11\rangle
-\frac{\sqrt{10}}{4}|\beta 11\rangle   \notag \\
&&+\frac{\sqrt{10}}{2}|\gamma 11\rangle ,  \label{met}
\end{eqnarray}
where the second member contains the following triplet states combining a
triplet state in bond $A$ with either a singlet, triplet or quadruplet in
bond $B$

\begin{eqnarray}
|\alpha 11\rangle  &=&|A:1,1\rangle |B:0,0\rangle ,  \notag \\
|\beta 11\rangle  &=&\frac{1}{\sqrt{2}}\left( |A:1,0\rangle |B:1,1\rangle
-|A:1,1\rangle |B:1,0\rangle \right) ,  \notag \\
|\gamma 11\rangle  &=&\sqrt{\frac{3}{5}}|A:1,-1\rangle |B:2,2\rangle 
-\sqrt{\frac{3}{10}}|A:1,0\rangle |B:2,1\rangle   \notag \\
&&-\sqrt{\frac{1}{10}}|A:1,1\rangle |B:2,0\rangle .  \label{triplets}
\end{eqnarray}

\begin{figure}[h!]
\begin{center}
\includegraphics*[width=1.0\columnwidth]{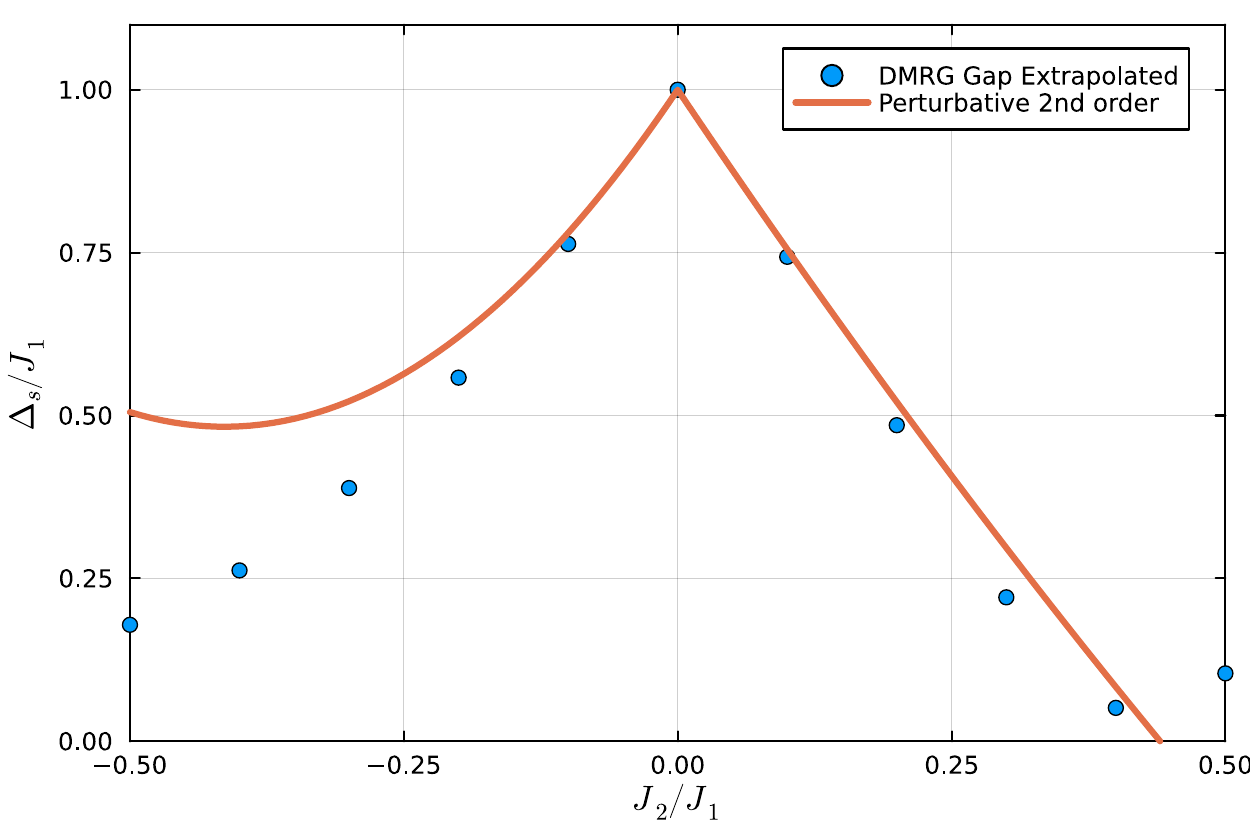}
\end{center}
\caption{Comparison of extrapolated DMRG results for the spin gap and the corresponding perturbative ones.}
\label{gapw}
\end{figure}

The gap to order 0 in $J_{2}$ is $\Delta _{0}=J_{1}$. The first term of Eq. (\ref{met}) is a hopping of the triplet to a nearest-neighbor bond. There is
a second-order contribution to this hopping coming from the process 
$|A:0,0\rangle |B:1,1\rangle \rightarrow |\beta 11\rangle \rightarrow |A:1,1\rangle
|B:0,0\rangle $ of magnitude $-5J_{2}^{2}/(16J_{1})$ (smaller than the first one). The energy of this state is minimized creating an excitation of wave
vector 0 or $\pi $ depending on the sign of $J_{2}$, and the gain of energy
is twice the effective hopping, leading to a contribution
\begin{equation}
\Delta _{1}=-\frac{5}{4}|J_{2}|(2+J_{2}/J_{1}).  \label{d1}
\end{equation}
The second-order processes that return to the initial state with either  $|\beta 11\rangle $ or $|\gamma 11\rangle $ as the intermediate state, and
including a factor 2 for contributions from the singlet bonds $A$ at the left
and the right of the triplet bond, contribute to the energy as
\begin{equation}
\Delta _{\beta }=\frac{-5J_{2}^{2}}{4J_{1}},\text{ }\Delta _{\gamma }=\frac{-5J_{2}^{2}}{3J_{1}}.  \label{d23}
\end{equation}
Finally to obtain the gap we have to add the difference with respect to the
singlet ground-state. Two contributions given by Eq. (\ref{me}) are lost:
\begin{equation}
\Delta _{s}=\frac{75J_{2}^{2}}{16J_{1}}  \label{ds}
\end{equation}
Adding all contributions we obtain for the gap
\begin{eqnarray}
\Delta  &=&J_{1}-\frac{5}{4}|J_{2}|(2+J_{2}/J_{1}) 
+\frac{85J_{2}^{2}}{48J_{1}}.
\label{gape}
\end{eqnarray}
Comparison between perturbative and DMRG results is shown 
in Fig. \ref{gapw}.
The agreement is good for small $J_2/J_1$, particularly 
for positive $J_2$.

The very existence of the strongly dimerized FM-AFM phase, uncovered by the DMRG exploration, makes the  perturbative results in the present Section a valuable tool to describe its features. Indeed the data in Fig.\ \ref{fig:u_0-vs-lambda} indicate that, close to the transition, $|J_2|/J_1$ is of the order of $10^{-2}$ with $J_2<0$. 

\section{About the presence of a spin Peierls effect for low spin-phonon coupling }
\label{low-lambda}

For low spin-phonon coupling one can use perturbative bosonization  to construct a low-energy effective theory  [\onlinecite{Schulz}]. 
It was found [\onlinecite{Affleck-Haldane,Affleck_1989,NAbos}] that for all half-odd integer spin there is a dimerization instability opening a spin gap $\Delta$ which should scale  as $\Delta \propto (\tilde{u}_0)^\nu$ with $\nu=2/3$. 
However, using perturbative bosonization [\onlinecite{Schulz}] the exponents have been shown to increase when 
$S> 1/2$. This may explain the smallness of the spin gap and the numerical difficulties to assess its existence.
In this context, numerical evidence for $S=1/2$, after hard efforts, was undoubtedly settled (see for instance [\onlinecite{undoubtedly}]). 
The spin $3/2$ case is far more challenging and, to our knowledge, 
there is neither a definitive confirmation yet nor recent efforts to elucidate the question.

\begin{figure}[ht!]
    \centering
    \includegraphics[width=1\linewidth]{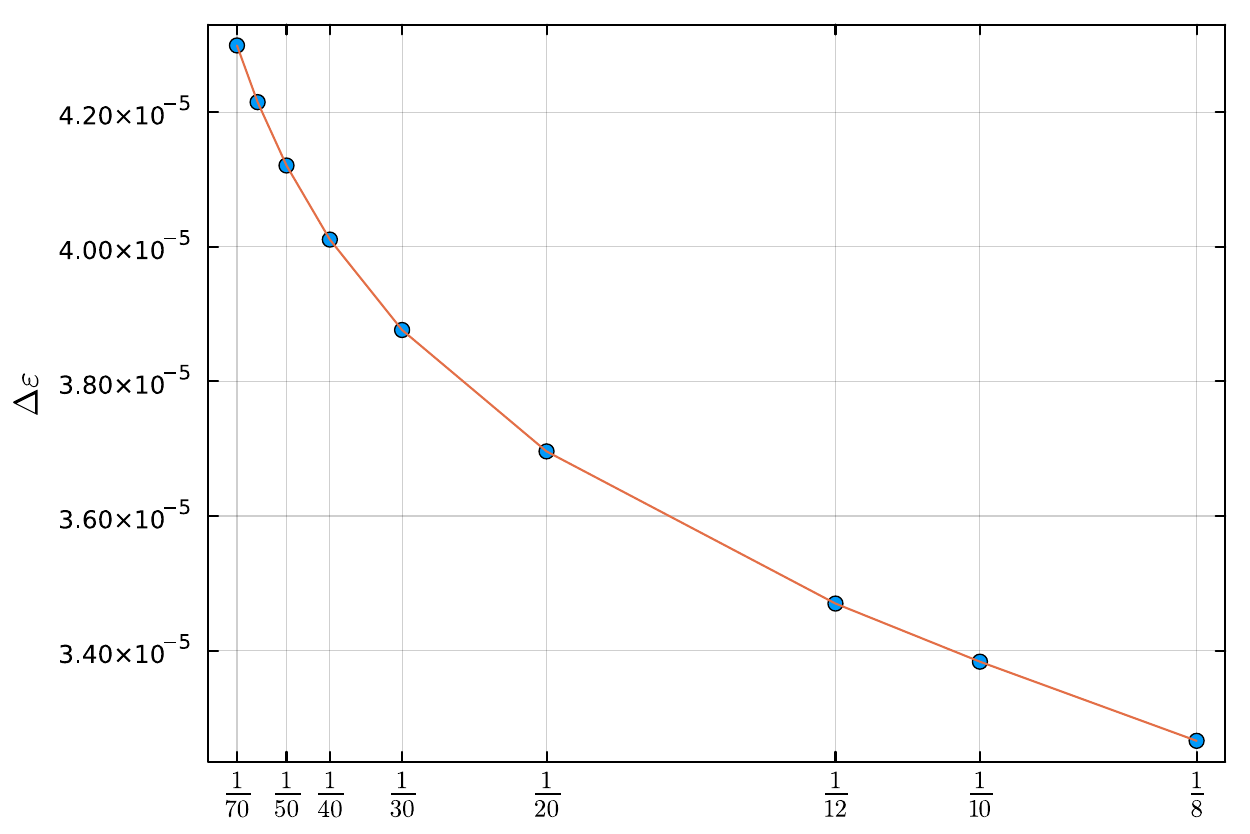}
    \caption{Size scaling of the gain in magnetic energy per site from alternating distortions with amplitude $\tilde{\delta}_0=0.005$, (update of the data in Fig.\ 1(c) in reference [\onlinecite{Guoetal}]). 
    The added data points with $12< N \leq 70$ confirm the convexity of the plot, supporting the existence of the spin gap.}
    \label{fig: update_Guo}
\end{figure}

Soon after the bosonization prediction, 
this case was analyzed in Ref.  [\onlinecite{Guoetal}] using exact  diagonalization for systems up to 12 sites; 
the finite size scaling behavior of the magnetic energy gain under dimerization (for a given amplitude) 
suggests that it overpasses the elastic energy cost for any spring constant $K$, 
leading to an unconditional (that is, for any 
$\lambda$) finite SP distortion  in the thermodynamic limit. 
Recall that the similar convexity of the plots in Fig.\ 1 of that reference for spin $S=1/2$ and spin $S=3/2$ is the key point in arguing the existence of the spin gap.
In the same spirit we have extended the analysis for larger chains  up to 70 sites, now available by DMRG. 
The results shown in Fig.\ \ref{fig: update_Guo}
confirm the current validity of their rationale.


This means that our curves in Fig.\ \ref{fig:energy-landscapes} should exhibit the first minimum close to, but not at $\tilde{u}_0 = 0$. However, due to the apparent smallness of the presumed distortion, a confident direct extrapolation to the thermodynamic limit is still unreachable within the available numerical precision. 

\section{Summary and discussion}

We have studied a magnetic chain coupled to the lattice for
the case in which the spin at each site is $S=3/2$. This value
is expected in transition-metal compounds due to the effect
of the Hund rules. The model, described in Section \ref{model},
contains a linear dependence of the exchange between
nearest-neighbor ions with their distance, and an elastic energy
loss when each ion is displaced from the equilibrium position.

In contrast to the $S=1/2$ case, we find a first-order transition
as a function of the dimensionless spin-phonon coupling $\lambda$
at a critical value $\lambda_c$. Our numerical results are consistent
with a vanishing or very small dimerization for $\lambda < \lambda_c$,
although from complementary studies we expect a small dimerization.

Instead, for $\lambda > \lambda_c$ there is a strong dimerization, with
large and small alternating exchange interactions $J_1$ and $J_2$.
The latter can even be negative within our model in this dimerized phase.
While this is not
expected for example in the superexchange for spins 1/2 in
the cuprates as the distance increases, a change in the sign
of the superexchange with structural parameters has been found
by theory and experiment in doped spin-1 $1T$-CrSe$_2$ [\onlinecite{freitas}].
The strongly dimerized phase is characterized by correlation functions
$\langle\mathbf{S}_{i} \cdot \mathbf{S}_{i+1}\rangle$ near
the ideal ones -15/4 for singlet dimers for the strong bonds, and
small, in general negative values for the weak bonds.
In particular, $J_2$ vanishes at the transition, 
rendering a system of perfectly decoupled singlet dimers.

When a magnetic field is included in the strongly dimerized phase, solitons separating the two possible dimerized phases (odd or even strong bonds) are formed. 

In the strongly dimerized phase, the large spin gap means the system does not gain as much energy from the applied magnetic field $h^z$
compared to the non-dimerized or weakly dimerized phase. This renders it possible to induce the transition with an external $h^z$, which could be relevant for applications. In many cases, distortions are associated with ferroelectricity and in this case, they could be significantly reduced by an external magnetic field. Conversely, an applied electric field could induce ferroelectricity and destroy the magnetic phase.
Another possibility is to tune $\lambda$ by striction or alloying.

\section*{Acknowledgements}

CJG is partially supported by CONICET (PIP 3220), Argentina.
AAA acknowledges financial support provided by PICT 2020A 03661 and PICT 2018-01546 of the Agencia I+D+i, Argentina. 
AOD is partially supported by UNR (80020190300213UR), Argentina.
GLR and DCC are partially supported by CONICET (PIP 1146), Argentina.


\begin{thebibliography}{99}

%
\bibitem{Cross} M.C. Cross and D.S. Fisher, 1
{\em A new theory of the spin-Peierls transition with special relevance to the experiments on TTFCuBDT},
Phys. Rev. B {\bf 19}, 402 (1979).

\bibitem{Feiguin_1997}  
A. E. Feiguin, J. A. Riera, A. Dobry, and H. A. Ceccatto, 
{\em Numerical study of the incommensurate phase in spin-Peierls systems},
Phys. Rev. B {\bf 56}, 14607 (1997).

\bibitem{Dobry-Riera_1995}  
J. Riera and A. Dobry,
{\em Magnetic susceptibility in the spin-Peierl	system CuGeO$_3$},
Phys. Rev. B {\bf 51}, 16098 (1995).

\bibitem{Penc_2004} 
K. Penc, N. Shannon, and H. Shiba, 
{\em Half-Magnetization Plateau Stabilized by Structural Distortion in the Antiferromagnetic
	Heisenberg Model on a Pyrochlore Lattice}
Phys. Rev. Lett. {\bf 93}, 197203 (2004).

\bibitem{Balents_2006}  
D.L. Bergman, R. Shindou, G.A. Fiete, and L. Balents,
{\em Models of degeneracy breaking in pyrochlore antiferromagnets}
Phys. Rev. B {\bf 74}, 134409 (2006).

\bibitem{Pouget} J.-P. Pouget, 
{\em Spin-Peierls, Spin-Ladder and Kondo Coupling in Weakly Localized Quasi-1D Molecular Systems: An Overview.} 
Magnetochemistry {\bf 9}, 57 (2023).

\bibitem{Hase}
M. Hase, I. Terasaki, and K. Uchinokura,
{\em Observation of the spin-Peierls transition in linear Cu$^{2+}$
(spin-1/2) chains in an inorganic compound CuGeO$_3$}
Phys. Rev. Lett. {\bf 70}, 3651 (1993).

\bibitem{Kiry}
V. Kiryukhin, B. Keimer, J. P. Hill, S. M. Coad, and D. M. Paul  {\em Synchrotron x-ray-scattering study of magnetic-field-induced transitions in Cu$_{1-x}$ (Zn, Ni)$_x$ Ge O$_3$} Phys. Rev. B, {\bf54}, 7269 (1996)


\bibitem{Temo}  C.J. Gazza, A.O. Dobry, D.C. Cabra, and T. Vekua, 
{\em Excitations with fractional spin less than 1/2 in frustrated magnetoelastic chains},
Phys. Rev. B {\bf 75}, 165104 (2007).

\bibitem{Guoetal} D. Guo, T. Kennedy, and S. Mazumdar,
{\em Spin-Peierls transitions in S$>$1/2 Heisenberg chains},
Phys. Rev. B {\bf 41}, 9592 (1990).

\bibitem{2D-SP} C.H. Aits, U. Löw, A. Klümper, and W. Weber
{\em Structural and magnetic instabilities of two-dimensional quantum spin systems}
Phys. Rev. B {\bf 74}, 014425 (2006);
Chenglong Jia, Jung Hoon Han
{\em Spin–lattice interaction effect in frustrated antiferromagnets}
Physica B {\bf 378–380}, 884 (2006). 

\bibitem{Haldane_1983}
F.D.M. Haldane,   
{\em Nonlinear Field Theory of Large-Spin Heisenberg Antiferromagnets: Semiclassically Quantized Solitons of the One-Dimensional Easy-Axis Néel State},
Phys. Rev. Lett. 50 1153 (1983).

\bibitem{Schulz} H. J. Schulz,
{\em Phase diagrams and correlation exponents for quantum spin chains of arbitrary spin quantum number}, 
Phys. Rev. B {\bf 34}, 6372 (1986).
\bibitem{Affleck-Haldane} 
I. Affleck and F.D.M. Haldane,
{\em Critical theory of quantum spin chains},
Phys. Rev. B {\bf 36}, 5291 (1987).

\bibitem{Affleck_1989} I. Affleck, D. Gepner, H. J. Schulz and   T. Ziman,
{\em Critical behaviour of spin-s Heisenberg antiferromagnetic chains: analytic and numerical results},
Journal of Physics A: Mathematical and General, {\bf 22}(5), 511 (1989).



\bibitem{NAbos} D. C. Cabra, P. Pujol, and C. von Reichenbach,
{\em Non-Abelian bosonization and Haldane’s conjecture},
Phys. Rev. B {\bf 58}, 65 (1998).

\bibitem{m1} 
{\em Magnetic order and magnetoelectric properties of R$_2$CoMnO$_6$6 (double) perovskites},
J. Blasco, J.L. Garc\'{\i}a-Muñoz, J. Garc\'{\i}a, G. Sub\'{\i}as, J. Stankiewicz,
J.A. Rodr\'{\i}guez-Velamazán, and C. Ritter, Phys. Rev. B \textbf{96}, 024409 (2017). 

\bibitem{m2} 
{\em Multiferroic behavior in the double-perovskite Lu$_2$MnCoO$_6$},
	S. Y\'a\~nez-Vilar, E.D. Mun, V.S. Zapf, B.G. Ueland,  J.S. Gardner, J.D. Thompson, J. Singleton,
	M. S\'anchez-And\'ujar, J. Mira, N. Biskup, M.A. Se\~nar\'{\i}s-Rodr\'{\i}guez, and C.D. Batista,
	Phys. Rev. B \textbf{84}, 134427 (2011).
	
\bibitem{m3} 
{\em Electric polarization observed in single crystals of multiferroic Lu$_2$MnCoO$_6$},
S. Chikara, J. Singleton, J. Bowlan, D.A. Yarotski, N. Lee, H.Y. Choi, Y.J. Choi, and V.S. Zapf,
Phys. Rev. B \textbf{93}, 180405(R) (2016).

\bibitem{m4} 
{\em Strong magnetoelectric coupling in mixed ferrimagnetic-multiferroic phases of a double perovskite},
M.K. Kim, J.Y. Moon, S.H. Oh, D.G. Oh, Y.J. Choi, and N. Lee,
Scientific Reports \textbf{9}, 5456 (2019).

\bibitem{m5}
{\em Magnetic domain wall induced ferroelectricity in double perovskites},
H.Y. Zhou, H.J. Zhao, W.Q. Zhang, and X.M. Chen,
Appl. Phys. Lett. \textbf{106}, 152901 (2015).

\bibitem{Pantografos}
D.C. Cabra, A.O. Dobry, C.J. Gazza, and G. L. Rossini,
{\em Microscopic model for magnetoelectric coupling through lattice distortions},
Phys. Rev. B  {\bf 100},  161111(R) (2019);
{\em Topological solitons and bulk polarization switch in collinear type-II multiferroics},
Phys. Rev. B  {\bf 103},  144421 (2021);
{\em Double frustration and magnetoelectroelastic excitations in collinear multiferroic materials},
Phys. Rev. B  {\bf 105}, 115103 (2022).


\bibitem{Pili}
L. Pili, R.A. Borzi, D.C. Cabra, and S.A. Grigera,
{\em Simple microscopic model for magneto-electric coupling in type-II antiferromagnetic multiferroics},
Phys. Rev. B  {\bf 105}, 094428 (2022).

\bibitem{ITensor} 
{\em The ITensor Software Library for Tensor Network Calculations},
M. Fishman, S.R. White and E. Miles Stoudenmire, SciPost Phys. Codebases, 4, 2022, URL https://scipost.org/10.21468/SciPostPhysCodeb.4

 
\bibitem{BPM-un-medio}
D. C. Cabra, A. Honecker, and P. Pujol,
{\em Magnetization plateaux in N-leg spin ladders},
Phys. Rev. B 58, 6241 (1998).

\bibitem{Poilblanc}
T. Vekua, D. C. Cabra, A. Dobry, C. Gazza, and D. Poilblanc,
{\em Magnetization Plateaus Induced by a Coupling to the Lattice},
Phys. Rev. Lett. {\bf 96}, 117205 (2006).

\bibitem{disclaimer} 
For large enough $\lambda$ we found a portion of the chain with a pronounced length contraction, 
ordered as in the present model, 
accompanied by a disordered portion with more separated ions to satisfy the total length constraint.

\bibitem{Vishwanath_2007}
F. Wang and A. Vishwanath,
{\em Spin phonon induced colinear order and magnetization plateaus in triangular and
kagome antiferromagnets. Applications to CuFeO$_2$},
Phys. Rev. Lett. {\bf 100}, 077201 (2008).


\bibitem{elsewhere}
The authors, in preparation.


\bibitem{undoubtedly}
M. Kumar, S. Ramasesha, D. Sen, and Z.G. Soos,
{\em Scaling exponents in spin 1/2 Heisenberg chains with dimerization and frustration studied with the density-matrix renormalization group},
Phys. Rev. B {\bf 75}, 052404 (2007).

\bibitem{freitas}
D. C. Freitas, M. N\'u\~nez, P. Strobel, A. Sulpice, R. Weht, A. A. Aligia, and M. N\'u\~nez-Regueiro,
{\em Antiferromagnetism and Ferromagnetism in Layered $1T$-CrSe$_2$ with V and Ti replacements},
Phys. Rev. B {\bf 87}, 014420 (2013).

\end{thebibliography}
\end{document}